# From Compressing Complexity to Accommodating Complexity: How AI Transforms Standardization and Individualization

**Li Li[1] ,Yu Cao[2]**
[1]Hefei No.62 Middle School, Hefei 230031,Anhui, China
[2]Anhui NARI ZT Electric Co., Ltd., Hefei 230031, Anhui,China

**Abstract**

Why do societies composed of individuals pursuing individuality repeatedly generate highly standardized systems? This paper argues that the answer lies in the evolution of information processing capacity. Artificial intelligence represents a historical transition in this capacity, enabling social systems to accommodate forms of complexity that previously had to be compressed. Industrial standardization was not merely a consequence of capital preference or power relations, but an institutional arrangement for maintaining the manageability of large-scale systems under limited information-processing capacity by reducing the variety of the controlled system. The fundamental change in the AI era lies in the expansion of information processing capacity across three dimensions: perception, computation, and execution. This expansion shifts personalized production from physical adaptation toward information-based adaptation and enables a transition from discrete to continuous objectification of difference. This paper proposes "cognitive fixed cost" as an analytical concept to describe how the upfront concentration of cognitive labor transforms the cost structure of personalized production. It further argues that standardization has not disappeared but has moved from explicit constraints at the product level to implicit generation rules embedded in infrastructures, shifting the central contradiction from "whether to have commonality" to "who controls commonality." The evolution of civilizational production logic is not a movement from commonality to individuality, but from compressing complexity to accommodating complexity.

## Chapter 1 The Historical Dilemma of Commonality and Individuality: Why Did Modern Civilization Choose Standardization?

### *1.1 The Problem*

Human beings are naturally diverse—in aesthetic preferences, needs, and experiences[1]. Yet modern society relies heavily on standard products, standard processes, and standard evaluations. The Model T came in only one color[2]; hamburgers taste the same around the world[3]; examinations measure different minds with the same test paper[4]. Why has a species that pursues individuality created a highly standardized civilization[5]?

In the industrial era, virtually every society that moved toward large-scale production—regardless of whether its capital structure was privately or state-owned, its political system democratic or authoritarian, its cultural tradition inclined toward individualism or collectivism—ultimately adopted standardized modes of production organization[6]. This cross-institutional universality suggests that beneath the phenomenon of standardization lie constraints more fundamental than capital logic[7], power structures[8], or cultural preferences[9].

*1.2 Existing Explanations and Their Limits*

Three mainstream explanations exist for industrial standardization.

The capital logic explanation holds that standardization pursues profit maximization—uniform specifications reduce production costs, simplify supply chain management, and expand market scale[7]. This explanation's contribution lies in revealing the functional coupling between standardization and capital accumulation, but its limitation is equally apparent: it cannot explain why reducing differences enhances efficiency. If personalization and high profit could be simultaneously achieved, capital would have no reason to refuse. Capital pursues profit, not standardization per se—standardization becomes the optimal path to profit only under specific conditions[10].

The power/discipline explanation holds that standardization is a technology of social control—uniform textbooks discipline thought, uniform factories discipline bodies, and uniform standards discipline consumers[8]. Foucault's theory of discipline provides a powerful analytical framework for this argument[8]. Yet this explanation invites a deeper follow-up question: why must control be exercised through standardization? If power could operate more effectively through differentiated means, there would be no reason for it to insist on standardization.

The efficiency explanation holds that standardization is inherently efficient—division of labor requires standard interfaces, scale requires standard units, coordination requires standard language. This explanation is closest to common sense, but it treats "standardization enhances efficiency" as a self-evident premise without interrogating the conditions under which efficiency holds. Smith's pin factory and Ford's assembly line were more efficient than artisan workshops not because "standardization" possesses some magic, but because the organizational technologies of the time could only coordinate large-scale production through standardization[10]. Efficiency is the result of standardization, not its cause—efficiency itself requires explanation.

The three explanations share an unexamined premise: that standardization is a function of efficiency. This paper asks: under what conditions does this function hold? Under what technological conditions does reducing differences enhance efficiency? When these conditions change, does the relationship between standardization and efficiency also change?

A more theoretically competitive question can clarify the explanatory boundaries of each theory: why did virtually every society moving toward large-scale production in the industrial era ultimately choose standardized modes of production organization? This cross-institutional universality serves as a touchstone for measuring each theory's explanatory power.

Transaction cost theory (Williamson, 1985) can explain firms' boundaries and choices of organizational form, but the cross-institutional universality of standardization exceeds its explanatory scope. Different institutional arrangements generate different transaction cost structures—if standardization were merely the result of transaction cost minimization, we should observe more diverse forms of production organization rather than universal convergence toward standardization[11]. Institutionalism (North, 1990) can explain institutional inertia but cannot explain why societies with different starting points and different paths ultimately converged on similar outcomes in the matter of standardization—path dependence logically leads to divergence, not convergence[12]. Power/discipline theory (Foucault, 1977) can explain how standardization serves social control but cannot explain why this particular form of control is so universally effective—the effectiveness of control depends on the feasibility of control, and feasibility is constrained by technological conditions[8].

The explanatory power of this paper's framework lies in this: the cross-institutional universality of standardization is rooted in the universal constraints on information processing capacity in the industrial era. Regardless of institutional arrangements, as long as information processing capacity is limited, large-scale production systems cannot respond to individual differences in real time—compressing differences becomes an inevitable choice for maintaining system manageability. The constraint of information processing capacity does not depend on institutional type but on technological conditions. This paper does not claim that information processing capacity is the sole cause of standardization; rather, it argues that it constitutes the boundary within which other factors operate. Capital can choose profit-maximizing paths within this boundary but cannot break through it to achieve real-time personalized production; power can design differentiated means of control within this boundary but cannot implement them without the requisite information infrastructure. Information processing capacity does not replace other explanations but delineates the scope of their validity.

*1.3 Core Hypothesis*

A society's capacity to accommodate differences depends on its information processing capacity.

This is not to say that information processing capacity is the sole cause of history. Capital, power, culture, institutions—these factors certainly play concrete roles in every historical period. But the operation of these factors is constrained by the information processing capacity of their era. This claim is not posited arbitrarily but builds upon existing foundations in information processing theory. Simon demonstrated that organizational decision-making is constrained by bounded rationality[13]; Ashby proposed the Law of Requisite Variety, arguing that a control system must possess variety matching that of the system it controls[14]. This paper extends these theories from the organizational level to the level of social production systems, arguing accordingly that the information processing capacity of each era constitutes the boundary of what production organizational forms can accommodate in terms of social differences.

Under the technological conditions of the industrial era, no matter how eagerly capital desired to satisfy individual needs, it lacked the information infrastructure for real-time perception, processing, and execution of massive personalized orders. Chandler's research on the coordination demands of modern enterprise management provides historical context for understanding the information processing constraints of the industrial era[15]. This historical context shows that large-scale production first needed to solve problems of organizational coordination and information processing. When an organization's information processing capacity is insufficient to respond to massive individual differences in real time, compressing differences becomes a practical path for reducing coordination costs. Under these technological conditions, capital's choices occur within the boundaries set by information processing capacity.

From this, a more general proposition can be advanced: the historical relationship between commonality and individuality fundamentally depends on society's information processing capacity. When information processing capacity is insufficient, commonality becomes a tool for compressing complexity — standardization is not a human preference but an institutional arrangement for maintaining the manageability of large-scale systems[14]. When information processing capacity breaks through, commonality transforms into infrastructure supporting individuality — standardization does not disappear but changes its mode of existence.

Manufacturing paradigms are moving from mass production through mass customization toward personalization[16]. This paper further argues that this transformation becomes possible in the AI era because the leap in information processing capacity extends the capacity to accommodate differences from fixed templates to an open space of individual parameters. Around this analytical framework, this paper further introduces such analytical concepts as information-based adaptation, cognitive fixed cost, and type implicitization to explain the specific changes in complexity governance in the AI era.

*1.4 Methodology and Conceptual Definitions*

This paper operationalizes "information processing capacity" into three sub-capacities: the capacity to perceive differences (acquiring and encoding individual-level difference information), the capacity to process differences (transforming difference information into production decisions), and the capacity to execute differences (adjusting physical production at low cost according to differentiated instructions). "Complexity" is operationalized as the sum of input differences, process differences, and output differences faced by the production system.

The paper employs "complexity" and "information processing capacity" as overarching concepts, under which correspond three interrelated technical concepts: Ashby's (1956) Variety—the number of possible states of the controlled system[14]; Galbraith's (1973) Information Processing Requirement—the amount of information a control system must process to cope with uncertainty[17]; and Simon's (1947) Complexity—the relative state that arises when information processing demands exceed the system's cognitive capacity[13]. The logical progression among the three is: the Variety of the controlled system determines the scale of information processing requirements; when information processing requirements exceed the system's information processing capacity, the system enters a state of Complexity. Throughout this paper, in different contexts where "complexity" is used, it follows this progressive logic: complexity is not an intrinsic property of the system but a relative relationship between the system's capacity and its environmental demands.

A methodological note on concept generalization: Ashby, Simon, and Galbraith discuss information processing capacity at the level of the single organization. This paper generalizes the concept to the scale of social production systems. The logic of this generalization is as follows: the production organizational forms of a given technological era are not determined by the choices of single organizations but are macro-patterns formed by countless organizations making adaptive choices under similar information processing constraints. When all organizations face the same ceiling on information processing capacity, their adaptive choices exhibit systematic convergence—this is the micro-foundation of cross-institutional universality.

This paper adopts the method of historical period comparison, using the history of technology and the history of production organization as primary empirical materials. It compares the different configurations of the relationship between information processing capacity and "commonality-individuality" across three periods—pre-industrial society, the industrial era, and the AI era—to test the validity of information processing capacity as an explanatory variable.

Finally, a methodological commitment applies throughout all chapters: this paper does not prove that AI necessarily brings freedom or liberation; it only explains why AI has, for the first time, changed society's capacity boundaries for processing differences. The production organizational forms of each era—whether the natural differences of the agricultural age, the standardization of the industrial age, or the scaled personalization of the AI age—are not objects of moral evaluation within this framework, but institutional arrangements formed under the constraints of information processing capacity. This non-moralizing analytical stance is the core marker distinguishing this paper from both technological utopianism and technological dystopianism.

## Chapter 2 Compressing Complexity: The Information Processing Boundaries of Industrial Civilization

*2.1 Differences in Pre-Industrial Society: Embedded in Experience, Not Yet Objectified*

Handcraft production in pre-industrial society was not devoid of difference. Ten tables made by the same carpenter would each differ due to variations in wood grain, tool condition, and the feel of the hand on a given day; blades forged in the same blacksmith's shop would differ due to fluctuations in furnace temperature and subtle variations in hammering force. But these differences are fundamentally different from what we today call "personalized production."

Distinguishing three types of difference helps clarify this distinction. Natural differences refer to uncontrollable variations caused by materials, environment, and tools; artisanal differences refer to inconsistencies produced by handcraft experience; design differences refer to customization actively undertaken for specific needs. Pre-industrial society was dominated by the first two. Although design differences existed—bespoke furniture for the nobility, altarpieces for the church, imperial porcelain for the court—they were always confined to a tiny elite stratum and likewise depended on the craftsman's direct experience. They lacked the conditions for scaled replication and were not the primary logic driving the operation of the production system.

These three types of difference share one fundamental feature: they are all embedded in the craftsman's bodily experience, concrete materials, and the context of making, without being extracted as information objects independent of the production process. The objectification of difference—extracting difference from concrete experience into recordable, transmittable, operable information—becomes economically rational only when the production system has the capacity to respond to differences systematically. Pre-industrial society's production was organized around workshops, its markets characterized by localization; the returns from objectifying differences did not cover the costs, so differences remained in experiential form. Industrial standardization itself precisely inaugurated the first step of difference objectification: it captured difference as finite discrete categories (S/M/L sizing, tolerance grades, product models), giving difference a symbolic form of existence for the first time, detached from the craftsman's body. But this discrete objectification simultaneously compressed the dimensions of difference—continuous individual variations were sorted into finite classificatory boxes; differences falling outside the boxes went unrecorded. The premise of personalized production is that difference can be objectified into operable information—this is a premise shared by both industrial standardization and AI-driven personalization, and one that pre-industrial society had not yet met. The leap from discrete objectification to continuous objectification will be developed in Chapter 3.

*2.2 The Complexity Bottleneck of Industrial Society*

Large-scale production systems face not a problem of efficiency but a problem of complexity management—this is also the fundamental point of divergence between this paper and efficiency theory: the key driver of standardization is not "faster, cheaper," but the fact that when a system's information processing capacity is insufficient to handle the variety it confronts, compressing variety becomes a prerequisite for maintaining the system's operation. When production scale expands from a workshop's dozens of items to a factory's millions, the quantity of information the system must process grows exponentially. Specifically, large-scale production requires three types of information processing capacity.

Perceiving differences—the capacity to acquire and encode the needs of all individuals. In the industrial era, perception relied on manual sampling and market research. Firms inferred the "average consumer" through questionnaires and sales data, rather than acquiring each consumer's actual needs. Sampling means information loss—differences are discarded at the perception stage.

Processing differences—the capacity to transform difference information into production decisions. In the industrial era, processing relied on pre-planned fixed workflows. Product specifications were determined at the design stage; process routes were frozen before production began. Any decision deviating from the standard path required escalation through management and human judgment. The production system lacked the capacity for real-time re-planning.

Executing differences—the capacity to adjust physical production at low cost according to differentiated instructions. In the industrial era, execution relied on specialized machinery—stamping dies, welding jigs, assembly-line cadences—all locked onto specific products. Changing a product meant changing tooling, re-debugging, and line downtime. Every differentiated response at the execution stage incurred significant friction costs.

All three capacities were strictly constrained under the technological conditions of the industrial era. Perception could not engage in full-scale collection; processing could not perform real-time reasoning; execution could not switch flexibly. Standardization was not a value preference but a strategy for maintaining system manageability under conditions of complexity overload. As Simon noted in his discussion of bounded rationality, when decision-makers face information processing demands exceeding their cognitive capacity, they do not seek optimal solutions but accept "good enough" satisficing solutions[13]. Industrial production systems faced an amplified version of the same dilemma: the system could not process every consumer's differences, so it accepted a simplified alternative—standard products.

This judgment is the key prior proposition of this paper: standardization requires explanation. It is not a natural expression of efficiency but an institutional arrangement for maintaining the operation of large-scale production systems under specific information processing constraints.

*2.3 Standardization as a Complexity Compression Mechanism*

Industrial-era standardization took two functional forms. The first is standardization at the level of production objects—unifying product specifications and compressing the dimensions of difference in final products. The second is standardization at the level of infrastructure—unifying interfaces and protocols, reducing coordination costs among system components, and releasing the combinatorial possibilities of the upper layers (as standard rail gauge released the expansion capacity of railway networks; standard voltage released the scale of the electrical appliance market). This paper addresses the first, which directly acts upon the degree of differentiation of final products and is the core mechanism of complexity compression. The second will be analyzed in Chapters 4 and 5 as the precursor to "commonality submerging into infrastructure."

Standardization does not eliminate individuality; it excludes the complexity that the system cannot process. The operation of this mechanism can be observed at three historical junctures, each acting upon a different level of the production process—motion, product, and inter-organizational interface—yet sharing the same underlying logic.

Taylorism decomposed complex craft skills into standardized sequences of motions. Taylor's classic study at Bethlehem Steel was not intended to "scientifically discover the one best way" but to eliminate individual differences in worker skill[18]. Each shoveling motion was decomposed into parametric combinations of angle, depth, and frequency; any behavior deviating from the standard motion was treated as efficiency loss. The tacit personal judgment embedded in craft—adjusting force according to material texture, adjusting rhythm according to fatigue—was stripped from the production process. Taylorism's core operation is not "acceleration" but "unification": excluding workers' bodily differences from the production system, making them interchangeable standard units. The essence of Taylorism is not making workers work harder but reducing the information processing burden on management by eliminating skill differences—when every worker's motions follow uniform norms, managers can supervise and schedule without understanding each worker's unique craft.

Fordism advanced standardization from the motion level to the product level. The case of the Ford Model T provides a precise observational window. Between 1908 and 1927, Ford produced over 15 million Model T vehicles, almost all in black. The popular explanation—that black paint dried fastest, suiting the assembly-line rhythm—is but a technical detail. The deeper reason is this: the assembly line of the time could stably respond to only one value of the variable "color." Adding color options meant: suppliers would need to provide multiple paint formulations rather than one; inventory would need to manage the matching relationships of parts corresponding to multiple colors rather than one-to-one correspondence; the line would need to clean spray guns and adjust parameters during color switches, increasing non-production time; management would need to track order volumes for each color to adjust production plans. With each addition of a color variable, the system's information processing load grew not linearly but in a chain reaction through the supply chain and management hierarchy. Under conditions where the primary information tools were manual scheduling, paper documents, and telephone coordination, the system could not bear this complexity. Ford's famous summary in his autobiography—"Any customer can have a car painted any color that he wants so long as it is black"[19]—was not expressing a personal preference but stating the boundary of the system's capacity.

The ISO standards system extended standardization from within single organizations to between organizations. When different factories, suppliers, and customers needed to collaborate, their independently defined quality standards constituted enormous information friction—Factory A's "qualified" was incompatible with Factory B's "qualified"; every transaction required renegotiation and re-inspection. ISO unified product tolerance grades, inspection procedures, and quality definitions into public standards, dramatically reducing the information costs of cross-organizational coordination[20]. The function performed by standardization here is: replacing countless private judgment standards with a single public discrete classification system.

Taylorism, Fordism, and ISO each addressed different segments of the production process, but they share a single logic: when the variety of the controlled system (worker skill, product specification, quality definition) exceeds the processing capacity of the control system (management, assembly line, supply chain), reducing the variety of the controlled object is the inevitable choice. This is the manifestation of Ashby's (1956) Law of Requisite Variety at the level of production organization. Ashby's law states that a control system can achieve effective control only when its own variety is no less than the variety of the system it controls[14]. The dilemma faced by industrial production systems is this: the variety of the controlled objects—millions of consumers, materials, processes—far exceeded the variety-accommodating capacity of the control system—the production organization based on manual management, mechanical control, and limited communication. The system therefore chose a reverse path: not to enhance its own complexity (which was infeasible under the technological conditions of the time) but to reduce the complexity of the controlled objects. Standardization is the institutionalized form of this reverse path.

The secret of industrial civilization's achievement of mass production lies not in the commonsense judgment that "standardization is more efficient" but in standardization's performance of a function previously left unnamed: complexity compression. By discretizing continuous individual differences into finite standard categories—compressing countless possible colors into "black," the infinite motion variations of craftsmen into "standard operating procedures," countless private quality judgments into "ISO tolerance grades"—it enabled the production system, with its extremely limited information processing capacity, to manage a material world whose scale far exceeded its cognitive capacity.

The standardization of industrial civilization is, in essence, an institutionalized response to the constraints of information processing capacity. Faced with complexity, a system has only two paths: enhance its own variety to match the variety of the controlled objects, or compress the variety of the controlled objects. Industrial civilization took the second path. Standardization is not a human preference but the only scalable strategy for maintaining the operation of large-scale systems when the first path is impassable.

---

*A system can only control the variety it has the capacity to match.*

## Chapter 3 Accommodating Complexity: How Artificial Intelligence Changes the Conditions of Personalization

### *3.1 The Three-Dimensional Leap in Information Processing Capacity*

Chapter 2 argued that industrial standardization was a strategy whereby systems maintained manageability by compressing the variety of controlled objects under conditions of limited information processing capacity. If this judgment holds, then the key question shifts from "is standardization efficient?" to "when information processing capacity undergoes qualitative change, how will the relationship between standardization and personalization be reconfigured?" This is precisely the core change of the AI era.

AI's transformation of information processing capacity is not a single-dimensional efficiency gain but a simultaneous leap across three dimensions: perception, computation, and execution.

Perception dimension: from manual sampling to full-scale collection. In the industrial era, firms inferred the "average consumer" through questionnaires, sales data, and market research—this is sampling logic; differences were smoothed out by statistical means at the perception stage. In the AI era, sensor networks, IoT devices, and real-time collection of user interaction data enable systems to acquire fine-grained information at the individual level. Perception no longer depends on pre-set classificatory frameworks but confronts raw difference directly.

Computation dimension: from fixed workflows to model inference. In the industrial era, once difference information was acquired, a second bottleneck emerged: how to process it. Management made decisions through pre-planned standardized workflows — product specifications were locked at the design stage, and process routes were fixed before production. Any demand deviating from the standard path required human judgment and approval through multiple managerial levels; decision costs grew linearly or even super-linearly with the number of differences. In the AI era, machine learning models complete the learning of large-scale difference patterns during training and perform parameterized decisions for individual cases within milliseconds during inference. Difference processing shifts from “re-analyze each time” to “learn once, invoke continuously.”

Execution dimension: from specialized machinery to dynamic parameter adjustment. The execution systems of the industrial era—stamping dies, welding jigs, assembly-line cadences—were locked onto specific products. Changing products meant changing physical tooling, re-debugging, and line downtime. Digital manufacturing systems—including programmable logic controllers, flexible machining cells, and digital-twin-driven production line configuration—enable equipment to change production behavior in real time according to differentiated instructions, switching freely within the parameter space without physical reconfiguration.

Although the change in the execution dimension is brought about by digital manufacturing systems, AI's breakthroughs in the perception and computation dimensions give the execution dimension an entirely new significance. In the industrial era, digital execution systems were not absent—CNC machine tools have been in use since the 1950s—but they faced the repetitive production of standardized products; their capacity for differentiation went uncalled. When AI enables systems to perceive individual differences and transform differences into production parameters in real time, digital execution systems for the first time genuinely face "differentiated execution instructions" rather than "repetitive execution of the same product." The leap across the three dimensions is not the sum of independent upgrades but a systemic synergy: AI expands the scope of perception and computation, thereby activating the differentiated response capacity that execution systems already possessed but had never fully utilized.

An important qualification applies to the changes in these three dimensions: this paper does not argue that AI has infinitely enhanced information processing capacity but that it has changed the cost structure and organizational boundaries of complexity processing. The three dimensions form an integrated technological system; breakthroughs in any single dimension alone are insufficient to change complexity-accommodating capacity — perceiving differences without being able to process them, or processing differences without being able to execute them, are merely partial improvements.

But the simultaneous leap across all three dimensions brings a qualitative change: for the first time, systems possess the information processing foundation to face large-scale differences without collapsing. This is not "working faster" but "being capable of processing complexity that was previously unprocessable." Industrial-era systems had to maintain operation by compressing differences—this is the core conclusion of Chapter 2. In the AI era, systems begin to acquire the capacity to accommodate differences without collapsing.

### *3.2 From Physical Adaptation to Information-Based Adaptation*

What has the leap in information processing capacity changed? The most fundamental change is this: the basic mode of complexity processing in personalized production has shifted from physical adaptation to information-based adaptation.

The core characteristic of traditional personalized production is physical adaptation. Each personalized requirement demanded a renewed physical cycle of redesign, retooling, and revalidation. A bespoke garment required the tailor to remeasure, re-pattern, and re-cut; a custom device required the engineer to recompute parameters, redraw blueprints, and produce a trial prototype. Every change in customer demand — color, size, configuration — triggered an iterative process in the physical world, with each step incurring incompressible time and material costs. Personalization costs grew linearly with the degree of differentiation; this physical constraint determined that large-scale personalization was economically infeasible.

In the AI era, the locus of processing personalized requirements has undergone a fundamental shift. Requirements are encoded as data inputs, transformed through model inference into parameter combinations, which then drive execution equipment to complete manufacturing. Design space search, solution simulation, process path planning — steps that previously required manual trial and error in the physical world — now occur in information space. The portion of complexity in personalized production that was originally borne by physical manufacturing has been transferred into computable information processing problems.

The internal mechanism of this transformation can be illuminated through a concept previewed in Chapter 2: the shift from discrete objectification to continuous objectification. As discussed in Chapter 2, industrial standardization captured difference through discrete objectification—sorting individual variations into finite classificatory boxes. What AI achieves is continuous objectification: differences are no longer forcibly assigned to preset categories but instead preserved as precise parameters of a single product. A consumer's body shape data is no longer "S or M or L" but a continuous vector composed of dozens of bodily dimensions; a product's color is no longer "black or white" but a precise chromatic value defined by spectral data. For the first time, difference enters the production system as itself — not as a label in a common-sense classification.

Siemens' Amberg factory provides a precise empirical observation window[21]. The factory's digital manufacturing system creates a digital twin for each product—before physical production, the product undergoes full-process simulation and parameter validation in information space. Different customers' differentiated requirements — interface configuration, firmware version, certification standards — complete compatibility checking and process path generation in information space before the product enters the physical production line. The individual differences of products have shifted from "physical trial and error" to "information simulation." This is not an acceleration of the manufacturing process but a transfer of the locus of complexity processing: complexity no longer piles up at the physical production stage — awaiting on-site engineer judgment, awaiting production line downtime for debugging—but is frontloaded into information space.

What requires precise clarification is that the core technological foundation of the Amberg factory is a digital manufacturing system (including digital twins, flexible production lines, and real-time data collection), not AI in the strict sense. What it demonstrates is the basic logic of information-based adaptation — the transfer of the locus of complexity processing from the physical world to information space—a logic already realized in digital manufacturing systems: digital twins map the differences of physical products into parameters in information space, and flexible production lines automatically adjust machining paths according to parametric differences. On this foundation, AI's further contribution lies in extending information-based adaptation from "rule-based" to "data-driven dynamic optimization" — for example, machine-learning-based quality prediction models that dynamically adjust process parameters based on real-time sensor data, and anomaly detection algorithms that identify defect patterns beyond the reach of fixed rules. Amberg demonstrates the foundation of information-based adaptation; AI extends the scope and intelligence of adaptation on top of that foundation. This is also why this paper takes "information processing capacity" rather than "artificial intelligence" as its core explanatory variable: the logic of information-based adaptation was already established in digital systems; AI extends it from finite rules to an open space of optimization.

What does this advance mean theoretically? Pine II (1993) proposed the concept of "mass customization" from an economic perspective, arguing that modular design and flexible manufacturing could achieve "personalized products at standardized cost" within a limited range[22]. Anderson (2006) argued from the long-tail effect that digital technology could reduce the distribution costs of differentiated products[23]. The shared limitation of both is that they treat personalization as an economic phenomenon — an optimization problem of cost structures and market scope. They explain how digitization reduces the cost of personalized products but do not answer a more fundamental question: why was personalization necessarily expensive before? Under what technological conditions was this "necessity" broken? This paper's advance lies in elevating this transformation from an economic phenomenon to a techno-philosophical proposition: what AI (and the digital technological system in which it is embedded) changes is not merely cost but the basic mode of complexity processing in personalized production. Traditional personalization means "facing each difference anew and going through another full physical-world cycle of design–manufacture–inspection"; information-based adaptation means "completing the analysis and mapping of differences once and for all in information space, with the physical world only responsible for execution." When the primary locus of complexity processing shifts from the physical world to information space, the long-standing contradiction between personalization and scale, for the first time, acquires the technological possibility of unification.

---

*From physical adaptation to information-based adaptation—personalized production no longer absorbs differences through physical iteration but transforms differences into a parametric mapping problem in information space. When the primary locus of complexity processing shifts from the physical world to information space, the long-standing contradiction between personalization and scale, for the first time, acquires the technological possibility of unification.*

*3.3 Cost Structure Transformation: "Cognitive Fixed Cost" as an Analytical Concept*

What has the shift from physical adaptation to information-based adaptation changed? A direct answer is "reduced costs," but this answer is too crude. A more precise judgment is: AI has not eliminated the costs of personalization but has changed the structure of costs. This section proposes "cognitive fixed cost" as an analytical concept to describe the core feature of this cost structure transformation.

Traditional personalized production was dominated by variable costs, the most critical of which was the variable cost of cognitive labor. Each time a new personalized demand was processed, the producer had to reinvest cognitive labor—understanding what the demand was, designing how to satisfy it, verifying whether the solution worked. The tailor remeasured each customer; the engineer recomputed parameters for each custom device; the designer re-researched and re-conceived for each project. These cognitive investments could not be reused across orders: the next time a new customer appeared, the tailor still had to measure from scratch; for the next customization project, the engineer still had to compute from scratch. In the cost structure of personalized production, cognitive labor grew linearly with the number of differentiated items, constituting the fundamental barrier to scaling.

In digital intelligent production systems, this cost structure has undergone a structural change. Digital manufacturing systems such as CAD, CAM, and ERP had already achieved the digital storage and reuse of design knowledge and process knowledge before AI—one product model could drive multiple production runs; one process path could be invoked across production lines. But the reuse scope of these systems was limited to fixed product templates and preset process paths: once customer demands exceeded the range covered by the templates, the system fell back to the mode of manual redesign. AI's contribution lies in extending the scope of cognitive reuse from fixed templates to an open parameter space. Machine learning models complete the one-time learning of large-scale difference patterns during training—learning the mapping between different body shapes and garment patterns, learning the correspondence between different configuration requirements and equipment parameters—and during inference can generate parametric solutions for each new personalized demand within the coverage of the training, with the marginal cognitive cost of each invocation approaching zero.

This paper characterizes this feature as "cognitive fixed cost"—an analytical concept for describing how a one-time cognitive investment in intelligent production systems transforms the cost characteristics of subsequent personalized invocations. This concept requires three strict qualifications.

First, it is not a fixed cost in the cost accounting sense (such as plant and equipment depreciation); it does not point to accounting items or investment decisions but to the structural frontloading of cognitive labor—transforming cognitive labor that previously had to be reinvested each time into a one-time learning process, such that subsequent invocations no longer incur equivalent cognitive costs.

Second, it is fundamentally distinct from Brynjolfsson and McAfee's (2014) "zero marginal cost of reproduction"[24]: the latter applies to digital reproductions, where the marginal cost of copying approaches zero; this paper addresses the personalized production of physical products—cognitive work is frontloaded for reuse, but material processing, assembly, and inspection still incur irreducible variable costs.

Third, it has clear conditions of applicability: it holds only when "cognitive work, once completed, can be reused." If each personalized demand involves entirely novel cognitive problems—beyond the coverage of model training, beyond the combinations expressible within the parameter space—then the system falls back to the traditional mode of cognitive labor, and the advantage of cognitive fixed cost disappears. AI does not unconditionally reduce the cost of personalization; it redraws the boundary between "reusable cognition" and "non-reusable cognition." The precise position of this boundary depends on the model's capacity range, but the boundary itself always exists.

From an economic perspective, this concept can be understood as extending Williamson's (1985) transaction cost theory[11]. Williamson argued that asset specificity—the degree to which an asset, once invested in a particular use, is difficult to redeploy—determines the choice of transaction governance structures. Cognitive fixed cost can be regarded as a distinctive form of specificity investment: the one-time investment in model training and system construction is highly specifically dedicated to personalized production in a given domain, yet once completed, it can support massive heterogeneous invocations. The uniqueness of this specificity lies in this—it is not locked to particular transaction partners or product types (as a die can only produce one kind of part) but is locked to a particular range of cognitive capacity (as a model can handle multiple types of differentiated requirements within a given domain). Traditional asset specificity creates "lock-in"—once invested, difficult to redirect; the specificity of cognitive fixed cost instead creates "release"—once invested, it releases a vast space of differentiated invocations. This reversal is the economic logic underlying the structural change in the cost of personalization.

---

*The structural transformation of personalization costs lies not in eliminating costs but in transforming cognitive labor from "reinvest each time" to "learn once, invoke continuously."*

## Chapter 4 From Type Replication to Instance Generation: An Ontological Transformation of Production Logic

*4.1 The Type-First Logic of Industrial Production*

Embedded within the operational logic of industrial production is a fundamental order: first define the type, then replicate instances. This order was not established or justified at the level of philosophical theory; rather, it was practically embedded into the underlying structure of the production system through repeated cycles of three operational stages: design, manufacturing, and inspection.

The design stage establishes the priority of the type. Product specifications — the diameter and thread pitch of an M6 screw, the tolerance range of a Grade A product, the structural parameters of a particular car model — are fully defined before production begins. The design document is not a description of any particular product but a collection of conditions that all products belonging to that model must satisfy. Before a single concrete product has been manufactured, the type already exists as an independent standard.

The manufacturing stage translates the type into a process of replication in the physical world. The core operation of batch production is ensuring that every instance approximates as closely as possible the specifications defined by the type — jigs position workpieces at design coordinates, dies shape materials into design forms, cadences lock operation times into design rhythms. Workpieces on the production line are interchangeable precisely because all workpieces point to the same type. The value of manufacturing lies not in creating differences but in eliminating differences to faithfully replicate the type. The inspection stage then enforces the type as a mechanism of exclusion: products whose deviation from the type definition falls within the tolerance range are qualified; those exceeding the range, regardless of direction or degree, are uniformly excluded from the circulation system.

The three stages—design pre-defining the type, manufacturing replicating according to the type, inspection using the type as a yardstick to exclude deviations—form a self-consistent closed loop. After this loop has been executed at large scale and over long duration, it sediments in the cognition of the participants in the production system — engineers, managers, workers, consumers—as a tacit presupposition: the type is what is "real"; the instance is merely a copy of the type. The pathway by which this presupposition forms is not philosophical reasoning but operational repetition. The physical architecture of the production line repeats "type first, instance second" every day; the institutional design of quality management repeats "deviation from type is defect" every day; the presentation format of product catalogs repeats "the product is defined by its model number" every day.

Standardization in industrial civilization is not merely a production organization strategy; it is simultaneously a mechanism for the generation of cognitive structures. A production logic that is repeatedly executed produces, in those who execute it, the corresponding ontological intuition. Type is prior to instance — this is not a proposition from philosophy textbooks but the tacit ontology that the production line, in its day-after-day operation, writes into the depths of industrial civilization.

*4.2 From Type Replication to Instance Generation*

AI-driven production systems change the relationship between type and instance. The production process no longer departs from a pre-defined fixed type to replicate copies; it departs directly from concrete demand data to generate unique instances. Personalized demands are encoded as parametric inputs, transformed through model inference into production instructions, which drive manufacturing equipment to output differentiated final products. The instance is no longer a copy of the type but becomes an independent production objective. Industrial production makes the type explicit — the type exists independently in the form of product specifications, and instances are replications of the type; AI production makes the type implicit— the type no longer appears as an independent product standard, and instances become the unfolding of generative rules.

What is needed here is a positive argument for why this process should be understood as "instance generation" rather than "dynamic type combination." A potential alternative hypothesis holds that LLMs internally still contain substantial type structures — tokens, embeddings, and clusters in latent space — and that the final output is merely a complex recombination of these types. From this perspective, "instances" are only surface phenomena; "types" remain the fundamental basis, and AI production has merely increased the flexibility of type recombination.

This alternative hypothesis confuses the essential difference between "type" and "generative rule." The essence of the type is to constrain the space of possibility—it excludes differences by stipulating the constraints the final product must satisfy. The M6 screw standard not only stipulates what a screw "ought to be" but more importantly excludes screws with a diameter of 5.9mm or 6.1mm from entering the category "M6." The mode of operation of the type is constraining: conform or not, no third possibility. The essence of the generative rule, by contrast, is to unfold the space of possibility—it does not stipulate what the final product "must be" but stipulates "how differences are produced." This distinction is not a verbal nuance but a difference between two fundamentally different operational logics.

Take GPT's text generation as an example. The Transformer architecture does not stipulate the concrete content of the output text — it does not preset templates or constraints for text categories such as "greeting," "apology," or "argument." It stipulates the probabilistic distribution mechanism of text generation: given the preceding token sequence, compute the conditional probability of the next token, and sample from that distribution to generate. Each sampling may take a different path, producing a different textual instance. This process is not selecting a variant from a "greeting type" to fill in; it is unfolding a unique trajectory across the continuous probability surface in token space. The reason GPT's output differs each time is not that the type pool has expanded but that the generative rule itself contains no type constraints.

The same logic applies to diffusion models' image generation: the model does not preset templates such as "cat" or "landscape" but learns the reverse diffusion process from noise to image. Each generation departs from different random noise, traverses different denoising paths, and produces different images—the differences arise from the unfolding of the same generative rule under different initial conditions, not from differences among type variants.

Tokens and embeddings in AI systems should not be understood as "types." They are not constraints on the final output but operational elements upon which the generative process depends for its unfolding. Tokens are discrete lexical units; embeddings are continuous vector representations; the two provide an operational space for generation but do not determine the concrete result of generation. When a production system no longer operates by "excluding products that do not conform to the type" but operates by "unfolding generative rules to produce differences," describing this system's logic as "instance generation" more accurately captures the fundamental turn in its operational mechanism than describing it as "dynamic type combination."

Two immediate qualifications are necessary. First, this transformation does not mean the disappearance of type structures. AI systems still contain multiple abstract structures—patterns extracted by attention heads, representations formed by feedforward layers, and clusters in parameter space. What genuinely changes is the relationship between types and final products. Types no longer explicitly define the final output in the form of independent product standards; instead, they transform from explicit product templates into implicit generative rules. What changes is the mode of existence and mechanism of operation of types, not whether types exist at all.

Second, at the technical level, AI enables instances to be generated without dependence on pre-defined explicit types, but such generation still occurs within the possibility space presupposed by the model. Model architecture, parameter capacity, and training data distribution together constitute an implicit generative boundary — within this boundary, differences can freely unfold; beyond it, certain differences cannot be generated. From the perspective of Deleuzian philosophy, this remains "permitted difference"—variation produced within the scope of existing generative rules—rather than absolute difference that breaks through all frameworks. Deleuze's ontology of difference functions in this paper as an "unfulfilled promise": it provides a critical yardstick for measuring the distance between AI's technical realization and the genuine liberation of difference. It is precisely this distance that forms the logical starting point of the critical reflection in Chapter 5. AI opens up a space of possibility for difference, but it does not eliminate the boundaries of that space.

### *4.3 The Role Transformation of Commonality: From Product to Infrastructure*

When type shifts from explicit template to implicit generative rule, the position and function of commonality within the production system change accordingly.

In the industrial era, commonality manifested itself as the final product — standardized product models were the direct outputs of production activities. The Model T, M6 screws, Grade A steel plates—commonality was the "thing" consumers purchased. In the AI era, commonality sinks into the foundational layer: the language understanding capabilities of large models, the resource scheduling mechanisms of operating systems, the manufacturing process knowledge bases of digital manufacturing platforms, and the transmission protocols of standardized data interfaces. These constitute the foundational capabilities, while the upper layer outputs infinitely differentiated product instances. Consumers no longer purchase "the same common product"; instead, they purchase "a unique instance generated from a common foundation."

The function of commonality changes as well. Industrial-era commonality performed the function of excluding differences—standardized products meant "everyone gets the same thing." AI-era commonality performs the function of supporting differences—a standardized foundation means "everyone shares the same set of differentiated generation capabilities." Commonality transforms from a framework that constrains individuals into a condition that empowers individuals. In the language established in Chapter 4: commonality transforms from a type-definer (stipulating what the product must be) into a generative-rule-provider (stipulating how differences are produced).

This transformation is conceptually simple—commonality becomes a means rather than an end — but its implications are far-reaching. It means that the zero-sum relationship between commonality and individuality that characterized the industrial era—more commonality means less individuality — is broken at the technical level. When commonality exists in the form of infrastructure, increases in commonality (more powerful foundation models, richer manufacturing platforms) instead bring about the expansion of the space of individuality (more diverse generative possibilities). Commonality no longer constrains differences; it empowers differences.

But this transformation simultaneously raises a new question: who controls this foundation? The retreat of commonality from the product level does not mean the disappearance of power relations but the transfer of power relations from the product level to the infrastructure level. This question is the core concern of Chapter 5.

---

*Type implicitization: the instance becomes the unfolding of generative rules. Commonality withdraws from the form of product into infrastructure, shifting from constraining differences to empowering differences.*

## Chapter 5 After the Release of Individuality: New Implicit Structures of Commonality in the AI Era

### *5.1 From Explicit Standardization to Implicit Standardization*

Industrial-era standardization was visible. Identical Model Ts rolled off the end of the assembly line; uniform school uniforms were worn on students' bodies; standardized examination papers were distributed to every desk — the mode of existence of standardization was empirically perceptible. Consumers knew that the product they bought was exactly the same as their neighbor's; this "sameness" itself was the most intuitive evidence of standardization.

AI-era standardization may be invisible. Algorithmic models, data architectures, and generative rules constitute the new form of standardization. Standardization shifts from the output level to the level of the generative mechanism, from "uniformity of product" to "uniformity of generative rules." When two users respectively pose questions to ChatGPT and receive different answers, superficially no standardization appears to be at work — the outputs are personalized. But these two different answers are produced within the same Transformer architecture, the same set of trained weights, the same RLHF alignment process. Difference occurs within a unified model framework — this is precisely the core meaning of implicit standardization.

The difference between explicit standardization and implicit standardization is not merely one of "visible" versus "invisible" but more fundamentally one of the object acted upon: explicit standardization acts on the final product—stipulating the diameter of the screw, the format of the exam — constraining "what is produced"; implicit standardization acts on the conditions of generation — stipulating model architecture, data formats, invocation protocols — constraining "how differences are produced."

Standardization shifts from an object that can be empirically perceived to a structure that requires technical analysis to reveal. The existence of explicit standardization is publicly visible and therefore can be identified, questioned, and resisted — consumers can see that their choices are the same as others' and can ask "why aren't there more options?" The risk of implicit standardization lies precisely in its invisibility: when everyone sees different output results, it is difficult to realize that these different outputs share the same underlying generative rules. The surface experience of personalization conceals the deep structure of standardization.

Foucault, in Discipline and Punish, described the shift of power technologies from visible displays of violence to invisible disciplinary technologies[8]. The shift from explicit standardization to implicit standardization structurally follows a similar logic—control no longer appears in the form of uniform products but operates through differentiated output forms. The difference is that what Foucault analyzed were intentionally designed mechanisms of power, whereas implicit standardization is more an unintended consequence of technical architecture.

*5.2 New Risks of Homogenization*

The concentration of foundation models may lead to a new type of homogenization — one that manifests not as convergence at the product level but as convergence in cognitive frameworks and preference-formation mechanisms. Three mechanisms merit attention.

The first mechanism is the continuous shaping of preference structures by algorithmic recommendation[25]. Recommendation systems do not directly tell users "what you should like"—this is precisely the subtlety distinguishing implicit standardization from explicit standardization. They gradually adjust recommendation strategies by continuously observing users' clicks, dwell time, and interaction behaviors, making recommended content increasingly fit the preferences users have already displayed. This process is formally responsive to individual differences but effectively locks users into the echo chamber of their existing preferences. Users' preferences are reinforced and narrowed in the process of being "responded to"; seemingly autonomous choices actually occur within a preset and progressively narrowing space of options.

The second mechanism is the embedding of value alignment within the boundaries of expression in foundation models[26]. Large language models undergo value alignment during training — methods such as RLHF embed human preference feedback into model weights, making models tend to generate "safe," "harmless," "norm-compliant" outputs. This alignment is operationally necessary, but its effect is to build value biases into the level of generative rules. What users receive are personalized outputs—different answers to different questions—but the generative space of these answers has already been pre-constrained by the alignment process. The scope of personalization occurs within the boundaries drawn by alignment.

The third mechanism is the large-scale stylistic convergence produced by generative AI. When millions of users use the same model to generate texts, images, and designs, the statistical features of model outputs—sentence structures, aesthetic tendencies, compositional logic—begin to form a recognizable "AI style" in cultural products. This style is not prescribed by any individual but emerges from the distribution of the model's training data. It redefines the baseline of "normal" and "innovative": normal expression increasingly approximates the statistical mean of model output; innovation means deviating from this mean. And the space of deviation—as argued in Chapter 4—is bounded by the model's generative limits.

These three mechanisms are not the recurrence of standardization in the traditional sense. They do not produce identical products; they do not enforce uniform behavior. They produce difference — but the systematic features of this difference (distribution, range, direction) are jointly determined by the underlying models and algorithms. This is a new type of structural convergence beneath the surface of individualized output.

*5.3 The Shift of the Core Contradiction: From "Whether to Have Commonality" to "Who Controls Commonality"*

When commonality submerges from the product level into the infrastructure level, the core contradiction in the discussion of commonality and individuality shifts accordingly.

The core contradiction of the industrial era was "commonality vs. individuality." Standardization and personalization stood in a zero-sum trade-off—more standardization meant less space for personalization. Consumers chose between buying standard products and ordering expensive customized ones, while firms traded off mass production against differentiated competition. The contradiction concentrated at the product level: commonality, embodied in product form, competed directly with individuality for the exact same space.

In the AI era, commonality and individuality no longer compete for the same space—they occupy different technological layers. Commonality resides at the foundation layer: large models, data infrastructures, and algorithmic frameworks; individuality emerges at the output layer: highly personalized recommendations, one-to-one generated content, and customized manufactured products. Personalization can occur at scale at the output layer, while commonality remains embedded at the foundation layer in a more concealed form. The contradiction therefore shifts from the product level to the infrastructure level—from "whether to have commonality" to "who defines the rules governing commonality at the foundation layer."

The AI-era discussion of commonality and individuality cannot remain at the level of "has AI released individuality?" "Release" itself is not a sufficient analytical concept—it does not answer under what conditions the release occurs, by whom its scope is delineated, or where its boundaries lie. These questions point toward the governance structure of underlying commonality. The openness of infrastructure—whether it is participatory, auditable, revisable—constitutes the decisive variable for the quality of the space of individuality. A closed black-box model can "release" individuality in the sense of outputting differentiated results, but it simultaneously delineates the generative boundaries of individuality, and the process of drawing these boundaries is invisible to users. Genuine release of individuality requires not only differentiated output at the output layer but that the rules of the foundation layer be transparent and amenable to participatory formulation.

Industrial-era power concentrated on "who can produce standard products"—the owners of the means of production set product standards. AI-era power shifts to "who can define underlying commonality"—the controllers of infrastructure set the conditions for the generation of differences. Lessig's (1999) proposition that "code is law" acquires new meaning here [27]: those who control the underlying code not only control how the system operates but also control the scope and structure of the differentiated space the system can produce. Commonality transforms from an explicit framework that constrains individuals into an implicit condition that empowers individuals—but empowerment is simultaneously constraint. The key question is not the presence or absence of commonality (commonality is technically unavoidable — any generative system requires underlying architecture) but the openness of commonality: who has the right to define the underlying rules, whether these rules are subject to public scrutiny, and whether alternative foundations are available for choice.

---

*The key question is not the presence or absence of commonality but the openness of commonality. Standardization retreats from the product level into the infrastructure level—more invisible, and more fundamental.*

## Chapter 6 Conclusion: The Historical Transformation of the Complexity Governance Paradigm

*6.1 Summary of the Three-Stage Evolution*

Taking information processing capacity as its explanatory axis, this paper has traced the evolution of the relationship between commonality and individuality across three historical stages. The following table presents their core differences:

| Era | Complexity State | Existential Form of Difference | Mode of Objectification | Resolution | Relationship Between Commonality and Individuality |
|---|---|---|---|---|---|
| Pre-Industrial Society | Unable to control differences | Embedded in experience, not objectified | None | Accept differences | Individuality as natural outcome, not scalable |
| Industrial Era | Unable to process differences | Standard types replace differences | Discrete objectification (finite classification) | Compress differences | Commonality prioritized, standardization as institutional arrangement |
| AI Era | Able to process differences | Differences objectified as data | Continuous objectification (individual parameters) | Accommodate differences | Individuality prioritized, commonality as support, commonality submerged into infrastructure |

In pre-industrial society, differences were "naturally occurring"—different wood, different touch, different furnace temperatures — but differences were not extracted as independent information objects. The craftsman's bodily experience was the sole carrier of difference; difference could not be recorded, transmitted, or systematically invoked independently of the concrete context of making. Individuality was a de facto existence, not a matter of design choice.

The industrial era objectified difference—size labels, tolerance grades, product models—giving difference, for the first time, a symbolic form of existence. But this objectification was discrete and compressed: continuous individual variations were sorted into finite classificatory boxes; differences falling outside the boxes did not exist within the production system. Standardization did not "eliminate" individuality but excluded the complexity the system could not process. Ashby's Law of Requisite Variety provides the cybernetic explanation for this mechanism: a system can only control the variety it has the capacity to match[14]. When information processing capacity is insufficient to accommodate the variety of the controlled objects, compressing the variety of the controlled objects is the inevitable choice.

The fundamental change of the AI era lies in this: the leap in information processing capacity—full-scale collection in the perception dimension, model inference in the computation dimension, dynamic parameter adjustment in the execution dimension—gives systems, for the first time, the capacity to accommodate differences without collapsing. Difference moves from discrete objectification to continuous objectification: individual differences are no longer forcibly assigned to preset types but enter the production system as precise parameters. The basic mode of complexity processing in personalized production shifts from physical adaptation to information-based adaptation — steps such as requirement analysis, solution generation, and process planning that previously required trial and error in the physical world are transferred into information space and completed once and for all. The long-standing contradiction between personalization and scale, for the first time, acquires the technological possibility of unification.

*6.2 Reaffirmation of the Core Proposition*

The historical relationship between commonality and individuality is fundamentally a function of society's capacity to process complexity.

When information processing capacity is insufficient, commonality becomes a tool for compressing complexity. Standardization is not a human preference, not a capitalist conspiracy, and not a ruse of power — although capital and power have indeed exploited and reinforced standardization — but an institutional arrangement for maintaining the manageability of large-scale systems under information processing constraints. That the same carpenter in pre-industrial society produced ten tables, each slightly different, does not signify personalized production; it signifies that differences could not be systematically controlled. That industrial society made a thousand workers perform the same standard motion does not stem from hostility toward individuality but from the ceiling of information processing capacity: under the technological conditions of the time, large-scale production could not be managed without compressing differences.

When information processing capacity breaks through, commonality transforms into infrastructure supporting individuality. Standardization has not disappeared but changed its mode of existence — from explicit constraints at the product level, submerging into implicit generative rules at the foundation level. The significance of the AI era is not the elimination of standardization but the fact that, for the first time in history under conditions of large-scale production, the capacity to accommodate differences has been extended from fixed templates to an open space of individual parameters, thereby rearranging the hierarchical relationship between commonality and individuality: commonality is no longer a force that constrains differences but a condition that empowers differences.

The evolution of civilizational production logic is not a movement from commonality to individuality but from compressing complexity to accommodating complexity. This core proposition is the first-order analytical framework of this paper — all the concepts discussed throughout, including information-based adaptation, cognitive fixed cost, type implicitization, and the submergence of commonality, are second-order explanatory mechanisms unfolded within this framework.

*6.3 Falsifiable Conditions and Scope of Applicability of the Theory*

Any theoretical framework must specify the conditions under which it could be falsified; otherwise it cannot be distinguished from untestable grand narratives. This paper's framework faces at least two conceivable empirical risks.

First, if there existed a society that, under the technological conditions of the industrial era, successfully achieved large-scale personalized production—maintaining scale efficiency without compressing differences—the core thesis of this paper would be weakened. A possible candidate is modular production, but modules themselves remain standardized; this merely adjusts the granularity of standardization from the whole machine to the component, rather than abolishing standardization. What would genuinely constitute a challenge is a form of large-scale production that entirely bypasses standardization.

Second, if the realization of large-scale personalized production in the AI era does not depend on enhanced information processing capacity but is achieved through purely institutional innovation, the validity of information processing capacity as the explanatory axis would be called into question. This paper's prediction is: institutional innovation can improve the efficiency of difference management but cannot substitute for information processing infrastructure—a system lacking digital perception, computation, and execution capacity, however its institutions are designed, cannot process massive personalized demands at scale efficiency.

On the scope of applicability: this paper's framework explains the relationship between commonality and individuality at the level of production organizational forms—that is, how production systems process and accommodate differences. It does not directly explain the expression of individuality in non-production domains such as culture, art, and intimate relationships, nor does it discuss why individuals desire differences (that is a question for the psychology of needs). It discusses what kinds of differences a social production system can accommodate under specific technological conditions, at what cost such differences can be accommodated, and how this capacity for accommodation becomes institutionalized into specific configurations of commonality and individuality. This boundary is not a defect of the framework but a condition of its analytical precision—a theory that can clearly state what it does and does not explain is more trustworthy than one that claims to explain everything.

### *6.4 Research Prospects*

This paper's framework takes the domain of production and manufacturing as its core empirical material; whether its explanatory power can be extended to domains such as service customization, content generation, and educational personalization awaits testing in subsequent research. In addition, two directions merit further development.

First, the empirical measurement of "cognitive fixed cost." This paper has argued at the conceptual level for the structural frontloading of cognitive labor and its impact on the cost structure of personalization but has not yet provided operationalized methods of measurement. Subsequent research could track, within specific industries, the proportional relationship between the one-time investment in model training and the marginal cognitive cost of subsequent personalized invocations, developing "cognitive fixed cost" from an analytical concept into a measurable empirical variable.

Second, the institutional design principles for "the openness of underlying commonality." Chapter 5 of this paper argued that the openness of commonality at the infrastructure level is the decisive variable for the quality of the space of individuality in the AI era, but it did not address the concrete institutional forms of openness. Subsequent research could explore possible paths for the governance of the foundation layer—open-source models, public data infrastructure, auditable algorithmic standards—and the actual impact of these paths on the space of personalization at the upper layer.

---

*The evolution of civilizational production logic is not a movement from commonality to individuality but from compressing complexity to accommodating complexity. The task of the AI era is not to eliminate commonality but to make commonality a participatory, auditable, and revisable public infrastructure.*

## References

1. Kahane G. Individuality as Difference[J]. Philosophy & Public Affairs, 2024, 52(4): 362-396. DOI: 10.1111/papa.12267.
2. Hounshell D. From the American system to mass production, 1800-1932: The development of manufacturing technology in the United States[M]. Jhu Press, 1984.
3. Ritzer G. The McDonaldization of society[M]//In the mind's eye. Routledge, 2021: 143-152.
4. Boake C. From the Binet–Simon to the Wechsler–Bellevue: Tracing the history of intelligence testing[J]. Journal of clinical and experimental neuropsychology, 2002, 24(3): 383-405. DOI: 10.1076/jcen.24.3.383.981.
5. Mumford L. Technics and civilization[M]. University of Chicago Press, 2010.
6. Kipping M. Consultancies, institutions and the diffusion of Taylorism in Britain, Germany and France, 1920s to 1950s[M]//Institutions and the Evolution of Modern Business. Routledge, 2014: 67-83.
7. Braverman H. Labor and monopoly capital: The degradation of work in the twentieth century[M]. nyu Press, 1998.
8. Foucault M. Discipline and punish: The birth of the prison[M]. Sheridan A, trans. New York: Vintage Books, 1977.
9. Hofstede G. The cultural relativity of organizational practices and theories[J]. Journal of international business studies, 1983, 14(2): 75-89. DOI: 10.1057/palgrave.jibs.8490867.
10. Farrell J, Saloner G. Standardization, compatibility, and innovation[J]. the RAND Journal of Economics, 1985: 70-83. DOI: 10.2307/2555589.
11. Williamson O E. The economic institutions of capitalism. Firms, markets, relational contracting[M]//Das Summa Summarum des Management: Die 25 wichtigsten Werke für Strategie, Führung und Veränderung. Wiesbaden: Gabler, 1985: 61-75. DOI: 10.1007/978-3-8349-9320-5_6.
12. NORTH D C. Institutions, institutional change and economic performance[M]. Cambridge: Cambridge University Press, 1990.
13. SIMON H A. Administrative behavior: A study of decision-making processes in administrative organization[M]. New York: Macmillan, 1947.
14. ASHBY W R. An introduction to cybernetics[M]. London: Chapman & Hall, 1956.
15. CHANDLER A D, Jr. The visible hand: The managerial revolution in American business[M]. Cambridge: Harvard University Press, 1977.
16. Hu S J. Evolving paradigms of manufacturing: From mass production to mass customization and personalization[J]. Procedia Cirp, 2013, 7: 3-8. DOI: 10.1016/j.procir.2013.05.002.
17. GALBRAITH J R. Designing complex organizations[M]. Reading, MA: Addison-Wesley, 1973.
18. Taylor F W. The principles of scientific management[M]. NuVision Publications, LLC, 1911.
19. Ford H, Crowther S. My life and work[M]. Binker North, 1922.
20. BRUNSSON N, JACOBSSON B. A world of standards[M]. Oxford: Oxford University Press, 2002.
21. Beitinger G, Monn P. Reinventing manufacturing: Beyond factory walls[J]. Journal of Supply Chain Management, Logistics and Procurement, 2022, 4(4): 354-368. DOI: 10.69554/KNUK7825.

22. PINE B J, II. Mass customization: The new frontier in business competition[M]. Boston: Harvard Business School Press, 1993.
23. ANDERSON C. The long tail: Why the future of business is selling less of more[M]. New York: Hyperion, 2006.
24. Brynjolfsson E, McAfee A. The second machine age: Work progress and prosperity in a time of brilliant technologies[M]. WW Norton & company, 2014.
25. Chaney A J B, Stewart B M, Engelhardt B E. How algorithmic confounding in recommendation systems increases homogeneity and decreases utility[C]//Proceedings of the 12th ACM conference on recommender systems. 2018: 224-232. DOI: 10.1145/3240323.3240370.
26. Casper S, Davies X, Shi C, et al. Open Problems and Fundamental Limitations of Reinforcement Learning from Human Feedback[J]. Transactions on Machine Learning Research, 2023. DOI: 10.3929/ethz-b-000651806.
27. Lessig L. Code: And other laws of cyberspace[M]. ReadHowYouWant. com, 2009.